\documentclass[12pt,preprint]{aastex}
\usepackage{natbib}
\usepackage{graphicx}
\shorttitle{A new target object}
\begin{document}
\title{A new target object for constraining annihilating dark matter}
\author{Man Ho Chan}
\affil{Department of Science and Environmental Studies, The Education University of Hong Kong}
\email{chanmh@eduhk.hk}

\begin{abstract}
In the past decade, gamma-ray observations and radio observations of our Milky Way and the Milky Way dwarf spheroidal satellite galaxies put very strong constraints on annihilation cross sections of dark matter. In this article, we suggest a new target object (NGC 2976) that can be used for constraining annihilating dark matter. The radio and x-ray data of NGC 2976 can put very tight constraints on the leptophilic channels of dark matter annihilation. The lower limits of dark matter mass annihilating via $e^+e^-$, $\mu^+\mu^-$ and $\tau^+\tau^-$ channels are 200 GeV, 130 GeV and 110 GeV respectively with the canonical thermal relic cross section. We suggest that this kind of large nearby dwarf galaxies with relatively high magnetic field can be good candidates for constraining annihilating dark matter in future analysis.
\end{abstract}


\section{Introduction}
In the past decade, gamma-ray observations and radio observations gave some stringent constraints for annihilating dark matter. For example, Fermi-LAT observations of the Milky Way center and the Milky Way dwarf spheroidal satellite (MW dSphs) galaxies give tight constraints on annihilation cross sections $<\sigma v>$ and dark matter mass $m$ for some annihilation channels \citep{Abazajian,Calore,Daylan,Abazajian2,Ackermann,Sameth,Li,Albert}. Also, radio observations of the Milky Way center put strong constraints on the annihilation cross sections of dark matter \citep{Bertone,Cholis,Cirelli2}. Generally speaking, our galaxy and the MW dSphs galaxies are the most important objects for constraining annihilating dark matter. It is because these objects are local or nearby objects so that the uncertainties of observations are generally smaller. Also, most of their properties including dark matter content are well-constrained. In particular, the MW dSphs galaxies are promising targets for detection due to their large dark matter content, low diffuse gamma-ray foregrounds and lack of conventional astrophysical gamma-ray production mechanisms \citep{Ackermann2}. Therefore, the constraints obtained are usually more stringent so that these objects are commonly believed to be the best targets for constraining annihilating dark matter.

Besides these objects, some recent studies use the data of M31 galaxy, M81 galaxy and some large nearby galaxy clusters (e.g. Coma, Fornax) to constrain annihilating dark matter \citep{Colafrancesco2,Egorov,Chan,Beck,Storm2}. These objects generally give similar or less stringent constraints compared with the Fermi-LAT observations of the Milky Way center and the MW dSphs galaxies. In this article, we explore a new target object (NGC 2976) and use its radio and x-ray data to constrain annihilating dark matter. We show that this object can give very strong constraints for annihilation cross sections, especially for three channels: $e^+e^-$, $\mu^+\mu^-$ and $\tau^+\tau^-$. 

\section{The x-ray constraints}
Generally speaking, an electron can increase a photon's energy from $E_0$ to $\sim \gamma^2E_0$ via inverse Compton scattering (ICS), where $\gamma$ is the Lorenz factor of the electron. If dark matter annihilates to give a large amount of high-energy positrons and electrons ($\sim$ GeV), these positrons and electrons would boost the energy of the cosmic microwave background (CMB) photons from $6 \times 10^{-4}$ eV to about 1 keV. Therefore, these photons can be detected by x-ray observations. 

However, this method is difficult to be used for normal galaxies and galaxy clusters because these objects usually emit strong x-ray radiation (due to hot gas). Unless we can accurately determine the thermal x-ray emission, the resulting constraints would be quite loose. For dwarf galaxies, this method can give much better constraints as the x-ray emission from dwarf galaxies is usually small (except those having AGN). Nevertheless, the size of a typical dwarf galaxy is small ($R \le 5$ kpc) so that the cooling rate of the high-energy electrons produced from dark matter annihilation is lower than their diffusion rate. Consequently, most of the high-energy electrons escape from the dwarf galaxy without losing most of their energy and the resulting x-ray signal is suppressed. For example, \citet{Colafrancesco,Jeltema} study the x-ray constraints for the local dwarf galaxies and find that the upper bounds of the annihilation cross sections are quite loose.

Fortunately, we realize a new target object, NGC 2976, which is a good candidate for applying this method. It is a relatively large nearby dwarf galaxy (linear size = 6 kpc, distance $d=3.5$ Mpc). Also, the total x-ray luminosity observed (0.3-8 keV) is $\sim 10^{36}$ erg s$^{-1}$, which is much lower than the other similar objects ($\ge 10^{38}$ erg s$^{-1}$ for others) \citep{Grier}. This relatively low x-ray luminosity can give tighter constraints for annihilating dark matter. Furthermore, the magnetic field strength of NGC 2976 is $B=6.6 \pm 1.8$ $\mu$G, which is relatively higher than the local group dwarf galaxies ($B=4.2 \pm 1.8$ $\mu$G) \citep{Drzazga} so that the cooling rate of high-energy electrons is higher than the diffusion rate. The cooling timescale for a 1 GeV electron in NGC 2976 is $t_c=1/b=7 \times 10^{15}$ s while the diffusion timescale is $t_d=R^2/D_0 \sim 10^{17}$ s, where $b \approx 1.4 \times 10^{-16}$ s$^{-1}$ is the total cooling rate, $R=2.7$ kpc is the isophotal radius of NGC 2976 \citep{Kennicutt}, and we have used a conservative diffusion coefficient $D_0=10^{27}$ cm$^2$ s$^{-1}$ \citep{Jeltema}. Therefore, we ensure that most of the high-energy positrons and electrons produced would loss most of their energy before escaping the galaxy. Since the diffusion process is not very important, the electron number density energy distribution function can be simply given by \citep{Storm}
\begin{equation}
\frac{dn_e}{dE}(\tilde{E})=\frac{<\sigma v> \rho^2}{2m^2b(\tilde{E})} \int_{\tilde{E}}^{m} \frac{dN'}{dE'}dE',
\end{equation}
where $\rho$ is the dark matter density profile, $dN'/dE'$ is the energy spectrum of the electrons produced from dark matter annihilation \citep{Cirelli} and $b(\tilde{E})$ is the total cooling rate, which is given by \citep{Colafrancesco2}
\begin{equation}
b(\tilde{E})=\left[0.25\tilde{E}^2+0.0254 \left( \frac{B}{\rm 1~\mu G} \right)^2 \tilde{E}^2 \right] \times 10^{-16}~{\rm GeV/s},
\end{equation}
with $\tilde{E}$ in GeV. Here, we neglect the Bremsstrahlung and Coulomb cooling as the thermal electron number density is very low in NGC 2976. 

The number of CMB photons scattered per second from original frequency $\nu_0$ to new frequency $\nu$ via ICS is given by
\begin{equation}
I(\nu)=\frac{3 \sigma_Tc}{16 \gamma^4} \frac{n(\nu_0)\nu}{\nu_0^2} \left[2 \nu \ln \left(\frac{\nu}{4\gamma^2\nu_0} \right)+\nu+4\gamma^2 \nu_0- \frac{\nu^2}{2\gamma^2 \nu_0} \right],
\end{equation}
where $\sigma_T$ is the Thomson cross section and $n(\nu_0)=170x^2/(e^x-1)$ cm$^{-3}$ is the number density of the CMB photons with frequency $\nu_0$, where $x=h\nu_0/kT_{\rm CMB}$. The total x-ray energy flux in the energy band $E_1$ to $E_2$ is given by
\begin{equation}
\Phi=2\times \frac{<\sigma v>J}{8\pi m^2} \int_{E_1}^{E_2}d(h\nu) \int_{m_e}^{m}\frac{Y(\tilde{E})}{b(\tilde{E})}d\tilde{E} \int_0^{\infty}I(\nu)dx,
\end{equation}
where
\begin{equation}
J=\int_{\Delta \Omega}d\Omega \int_{\rm los} \rho^2 ds
\end{equation}
is called the J-factor and
\begin{equation}
Y(\tilde{E})=\int_{\tilde{E}}^m \frac{dN'}{dE'}dE'.
\end{equation}
The dark matter density profile $\rho$ for NGC 2976 can be modeled by
\begin{equation}
\rho=\rho_0\left[1+\left(\frac{r}{r_c} \right)^2 \right]^{-1},
\end{equation}
where $\rho_0=0.198M_{\odot}$ pc$^{-3}$ and $r_c=1$ kpc \citep{Adams}. In addition, the substructures in NGC 2976 can greatly enhance the annihilation rate. By using a conservative model of substructure contributions \citep{Moline}, the substructure boost factor is about $B_f=4.44$. By considering the dark matter contribution within the isophotal radius $R=2.7$ kpc, we get $\log (J/\rm GeV^2~cm^{-5})=17.6$. There is another dark matter profile $\rho=\rho_0(r/\rm 1~pc)^{-0.235}$ with $\rho_0=0.260M_{\odot}$ pc$^{-3}$ which can produce good fit to the kinematic data of NGC 2976 \citep{Adams}. The corresponding J-factor for this dark matter profile is $\log (J/\rm GeV^2~cm^{-5})=17.5$. Therefore, the systematic uncertainty of the J-factor is about 30\%. In the following, since the uncertainty is not very large, we will use the dark matter profile in Eq.~(7) to perform the analysis. The effect of this uncertainty will be discussed later.

The total x-ray flux observed (0.3-8 keV) for NGC 2976 is $\Phi=(0.42 \pm 0.17) \times 10^{-14}$ erg cm$^{-2}$ s$^{-1}$ \citep{Grier}. By assuming that the observed x-ray flux originates from ICS due to dark matter annihilation only, we can obtain the upper limits of the annihilation cross sections for different channels (see Fig.~1). Here, we can see that the observed x-ray range is not very good for constraining annihilating dark matter. As we can see from Fig.~1, the x-ray constraints are close to the upper bounds obtained by Fermi-LAT observations for $e^+e^-$ and $\mu^+\mu^-$ channels only \citep{Ackermann}. For the thermal relic cross section $<\sigma v>=2.2 \times 10^{-26}$ cm$^3$ s$^{-1}$ \citep{Steigman}, the minimum allowed $m$ for the $e^+e^-$ channel is 8 GeV, which is slightly tighter than the Fermi-LAT limit for the Milky Way's dwarf galaxies \citep{Sameth}. If we can have better x-ray data or more hard x-ray data, the corresponding constraints would be much tighter.

Note that we have only included the CMB photons in our calculations. In fact, there are other radiation fields in the infra-red and visible light bands which can also contribute to the x-ray flux via ICS. Nevertheless, the contribution of other radiation fields in NGC 2976 is small and most of the resulting photons via ICS are in MeV or above bands. Therefore, our results would not be significantly affected by other radiation fields.

\begin{figure}
\vskip 10mm
 \includegraphics[width=140mm]{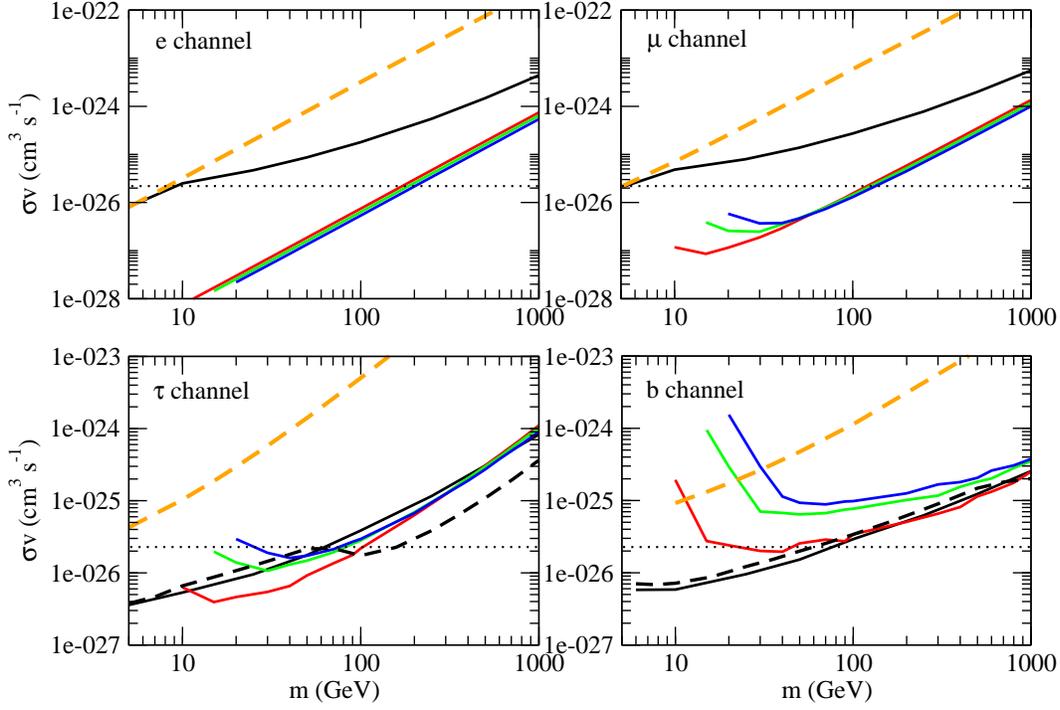}
 \caption{The upper limits of the annihilation cross sections for four annihilation channels. The red, green and blue solid lines represent the upper limits for the radio constraints (red: $\nu=1.43$ GHz; green: $\nu=4.85$ GHz; blue: $\nu=8.35$ GHz). The orange dashed lines represent the upper limits for the x-ray constraints. The black solid lines represent the gamma-ray observations of MW dSphs galaxies with Fermi-LAT (with J-factor uncertainties) \citep{Ackermann}. The black dashed lines represent the gamma-ray observations of recently discovered Milky Way satellites with Fermi-LAT (only for $\tau^+\tau^-$ and $b\bar{b}$ channels) \citep{Albert}. The dotted lines represent the canonical thermal relic cross section for annihilating dark matter \citep{Steigman}.}
\vskip 10mm
\end{figure}

\section{The radio constraints}
If we assume that all the radio radiation originates from synchrotron radiation of the electron and positron pairs produced by dark matter annihilation, the observed upper limit of the total radio flux can be used to constrain the cross sections of dark matter annihilation. As mentioned above, since the diffusion term can be neglected, the injected spectrum of the electron and positron pairs is proportional to the source spectrum \citep{Storm}. By using the monochromatic approximation (the radio emissivity is mainly determined by the peak radio frequency), the total synchrotron radiation energy flux of the electron and positron pairs produced by dark matter annihilation is given by \citep{Bertone,Profumo}:
\begin{equation}
S \approx \frac{1}{4 \pi d^2} \left[ \frac{9 \sqrt{3}<\sigma v>}{2m^2\tilde{b}} \int_0^R 4 \pi r^2 \rho^2EY(E)dr \right],
\end{equation}
where $E=0.43(\nu/{\rm GHz})^{1/2}(B/{\rm mG})^{-1/2}$ GeV and $\tilde{b} \approx 1.18$ is a correction factor if we include the cooling of ICS. 

Latest radio observations with three different frequencies $\nu=1.43$ GHz, $\nu=4.85$ GHz and $\nu=8.35$ GHz obtain $2-\sigma$ upper bounds of radio fluxes $S \le 1.02 \times 10^{-15}$ erg cm$^{-2}$ s$^{-1}$, $S \le 1.59 \times 10^{-15}$ erg cm$^{-2}$ s$^{-1}$ and $S \le 1.76 \times 10^{-15}$ erg cm$^{-2}$ s$^{-1}$ respectively \citep{Drzazga}. By using Eq.~(8), we obtain the corresponding upper limits of annihilation cross sections for four popular channels: $e^+e^-$, $\mu^+\mu^-$, $\tau^+\tau^-$ and $b\bar{b}$ (see Fig.~1). For the canonical thermal relic cross section, the minimum allowed $m$ for the $e^+e^-$, $\mu^+\mu^-$ and $\tau^+\tau^-$ are 200 GeV, 130 GeV and 110 GeV respectively. For the $b\bar{b}$ channel, the radio constraints just marginally disfavor the range $20$ GeV $\le m \le 50$ GeV. If we take the systematic uncertainty of the J-factor into account, the minimum allowed $m$ would decrease by about 15\%. Generally speaking, the radio constraints of NGC 2976 are tighter than the Fermi-LAT constraints \citep{Ackermann,Albert}, except for the $\tau^+\tau^-$ channel with $m \ge 100$ GeV and $b\bar{b}$ channel with $m \le 200$ GeV. For the $e^+e^-$ and $\mu^+\mu^-$ channels, the upper limits of annihilation cross sections are at least an order of magnitude tighter than the Fermi-LAT constraints \citep{Ackermann}. Therefore, we can see that NGC 2976 is a very good candidate for constraining annihilating dark matter. 

\section{Discussion}
In this article, we discuss a new target object, NGC 2976, for constraining annihilating dark matter. Generally speaking, nearby dwarf galaxies are good objects for constraining annihilating dark matter because they are rich in dark matter content and the effect of baryons is relatively small. However, since most dwarf galaxies are small and the magnetic fields are weak ($B<5$ $\mu$G), the diffusion of high-energy electrons and positrons would be quite efficient. Most of the electrons and positrons would escape from the dwarf galaxies without losing all of their energy. This suppresses the signals (both radio and x-ray) detected so that the constraints obtained would not be very tight. Nevertheless, this problem would be alleviated if a dwarf galaxy contains a relatively high magnetic field ($B \ge 5$ $\mu$G) and its size is large. The high magnetic field would greatly enhance the cooling rate so that most electrons and positrons would lose their energy within their stopping distance. 

As we mentioned, NGC 2976 has a high magnetic field $B=6.6 \pm 1.8$ $\mu$G and it is a relatively large dwarf galaxy (linear size = 6 kpc) so that the cooling timescale is much shorter than the diffusion timescale. This can maximize the x-ray and radio fluxes due to dark matter annihilation. The other advantage of using NGC 2976 is that it has tight upper bounds of x-ray flux and radio fluxes. These features suggest that NGC 2976 is a very good candidate for constraining annihilating dark matter, especially for the leptophilic channels. Further observations of NGC 2976 in radio wavelengths and x-ray bands can definitely push the upper bounds of cross sections to a much tighter level.  

In our analyses, we can see that the x-ray constraints of NGC 2976 are not good enough to give tighter constraints compared with the Fermi-LAT gamma-ray constraints. The limitations of using x-ray constraints are still quite large, unless we can fully identify the contribution of non-thermal x-ray emission. Nevertheless, the radio constraints can give much tighter constraints, which is complementary to the gamma-ray constraints. These constraints can rule out some existing models (via $e^+e^-$, $\mu^+\mu^-$ or $\tau^+\tau^-$) of dark matter interpretation of the excess gamma-ray and positrons in our galaxy \citep{Calore,Boudaud}. 

\begin{acknowledgements}
This work is supported by a grant from The Education University of Hong Kong (Project No.:RG4/2016-2017R).
\end{acknowledgements}

\end{document}